\documentclass[sensors,article,accept,moreauthors,pdftex,10pt,a4paper]{mdpi} 

\firstpage{1} 
\pubvolume{xx}
\issuenum{1}
\articlenumber{1}
\pubyear{2018}
\copyrightyear{2018}
\externaleditor{Academic Editor: name}
\history{Received: date; Accepted: date; Published: date}

\Title{Optical oxygen sensing with artificial intelligence}

\Author{Umberto Michelucci$^{1}$, Michael Baumgartner$^{2}$, Francesca Venturini$^{1,2}$*}

\AuthorNames{Umberto Michelucci, Michael Baumgarten, Francesca Venturini}

\address{%
$^{1}$ \quad TOELT LLC; Birchlenstr. 25, 8600 Dübendorf, Switzerland\\
$^{2}$ \quad Institute of Applied Mathematics and Physics, Zurich University of Applied Sciences,
Technikumstrasse 9, 8401 Winterthur, Switzerland}

\corres{Correspondence: francesca.venturini@zhaw.ch, umberto.michelucci@toelt.ai}

\abstract{Luminescence-based sensors for measuring oxygen concentration are widely used both in industry and research due to the practical advantages and sensitivity of this type of sensing. The measuring principle is the luminescence quenching by oxygen molecules, which results in a change of the luminescence decay time and intensity. In the classical approach, this change is related to an oxygen concentration using the Stern–Volmer equation. This equation, which in most of the cases is non-linear, is parametrized through device-specific constants. Therefore, to determine these parameters every sensor needs to be precisely calibrated at one or more known concentrations. This work explores an entirely new artificial intelligence approach and demonstrates the feasibility of oxygen sensing through machine learning.
The specifically developed neural network learns very efficiently to relate the input quantities to the oxygen concentration. The results show a mean deviation of the predicted from the measured concentration of 0.5 \% air, comparable to many commercial and low-cost sensors.
Since the network was trained using synthetically generated data, the accuracy of the model predictions is limited by the ability of the generated data to describe the measured data, opening up future possibilities for significant improvement by using a large number of experimental measurements for training. 
The approach described in this work demonstrates the applicability of artificial intelligence to sensing technology and paves the road for the next generation of sensors.}

\begin{document}

\section{Introduction}
\label{Introduction}

The determination of oxygen partial pressure is of great interest in numerous areas including medicine, biotechnology, and chemistry. Since oxygen or better dioxygen, plays an important role in many processes, applications range from biomedical imaging, packaging, environmental monitoring, process control, and chemical industry, to mention only a few.

There are different methods used to determine oxygen concentration depending on the application. Among these, optical methods are particularly attractive because they do not consume oxygen, and are therefore reversible, have a fast response time, and allow a good precision and accuracy. Additionally optical sensors can be manufactured with small sizes, mounted on a fiber and allow therefore both remote and in-situ measurements. An extensive review of optical methods for oxygen sensing and imaging can be found in \cite{Wang2014}. 

A well-known approach, successfully industrialized for several years, is based on the quenching of luminescence by the oxygen molecules \cite{Wolfbeis2004}. 
A dye molecule, also called here indicator, is embedded in a matrix permeable for oxygen. Its luminescence is quenched due to the dynamical collisions with oxygen. This process leads to a reduction by an amount which depends on the oxygen concentration of both the intensity and decay time of the luminescence  \cite{Demas1999,Lakowicz2006,Quaranta2012}. An overview of commercially available oxygen sensors based on luminescence quenching can be found in \cite{Wolfbeis2015}.

Sensors based on this principle rely on approximate empirical models to parametrize the dependence of the sensing quantity (e.g., intensity or decay time) on influencing factors. Additional to oxygen concentration, for example, temperature strongly influences the measurement, since both the luminescence and the quenching phenomena are temperature dependent. Other factors impacting the sensing quantity may be sensor specific and affected by the fabrication characteristics. These include, for example, the quenching rate constant of the indicator and the solubility of oxygen in the matrix which serves as a solvent for it. As a result, the characteristic is specific to the system used  \cite{Xu1994,Draxler1995,Hartmann1996,Mills1998,Badocco2008,Dini2011}.

As described in detail in the next section, the conventional approach consists in using a complex and frequently empirical multi-parametric model to calculate the oxygen concentration from the sensing quantities. Additionally, hardware-related factors, due for example to mechanical and optoelectronic tolerances, must be considered when implementing a luminescence-quenching scheme for commercial applications. 
The complexity of the model depends on the accuracy and working range which it is sought for. To reach the accuracy of commercial sensors \cite{Wolfbeis2015}, not only the dependence of the oxygen concentration from the measured quantity has to be approximated analytically, but also the non-linear dependence of these parameters from their influencing factors, e.g., temperature, has to be included. This impacts on the requirements for the electronics performing the calculation. Frequently, a compromise between accuracy, choice of the implementation microelectronic platform, costs and rapidity of the measurements has to be made.

In this work, the application of a machine learning algorithm to luminescence quenching is explored for the first time. To the best of the author's knowledge, machine learning was applied to time-resolved luminescence data \cite{Dordevic2018}, but never to phase-fluorimetry based luminescence sensing. The proposed novel approach consists in using a neural network which learns to relate the sensed quantities to an oxygen concentration value. The work explores but not exhausts the possibilities of this approach. For example, the training of the neural network is performed here with an artificially-created training dataset due to the limited number of measurements available. Additionally, temperature, which needs to be considered in luminescence measurements, was kept constant here but could be a parameter predicted by the neural network.

The best-performing network is identified through the study of different architectures and hyperparameter tuning. The trained network is then applied to experimental data, to check its generalization capacity when applied to unseen data. The analysis of the error in the prediction of the oxygen concentrations shows that is due to the synthetically training data, which are calculated with an approximate analytical model. Using for the training dataset experimental data will therefore reduce the absolute error that can be achieved with this approach. The present work shows that, on one hand the concept as implemented already achieves results comparable with many high-concentration (up to air concentration) commercial sensors and compact sensors based on the classical approaches \cite{Wolfbeis2015,Chu2016}, on the other hand, has still potential for improvement of the accuracy. The advantage of the proposed new approach is that the analytical model, and therefore its complexity, do not play any role. As long as enough data are available to train the neural network, the latter will be able to the learn dependence of the oxygen concentration from all the parameters.

The paper is organized as  follows: Section \ref{Theory_Lum} describes analytical models for oxygen luminescence quenching; Section \ref{Experimental} illustrates the experimental setup used for the experiments; the neural network and its tuning are discussed in Section \ref{Neuro}; the results are discussed in Section \ref{Results}.

\section{Theoretical model for the luminescence quenching}
\label{Theory_Lum}

Oxygen-quenching luminescence sensors are based on the decrease of the luminescence intensity and decay time of the indicator as a function of O$_2$ concentration. In the presence of molecular oxygen, the luminescence of the indicator is quenched because of the radiationless deactivation process due to the interaction of the indicator with molecular oxygen (collisional quenching). In the case of homogeneous media characterized by an intensity decay which is a single exponential, the decrease in intensity and lifetime are both described by the Stern-Volmer (SV) equation \cite{Lakowicz2006,Quaranta2012}
\begin{equation}
\frac{I_0}{I}=\frac{\tau_0}{\tau}=1+K_{SV} \cdot \left[O_2\right]
\label{SVe}
\end{equation}
where $I_0$ and $I$, respectively, are the luminescence intensities in the absence and presence of oxygen, $\tau_0$ and $\tau$ the decay times in the absence and presence of oxygen, $K_{SV}$ the Stern–Volmer constant and $\left[O_2\right]$ indicates the oxygen concentration.

In many practical applications, the indicator is embedded in a matrix or substrate, frequently a polymer. In this case, the SV curve $I_0 / I (\left[O_2\right])$ deviates from the linear behavior of equation (\ref{SVe}) \cite{Wang2014}. The deviation is attributed, for example, to heterogeneities of the micro-environment of the luminescent indicator, or the presence of static quenching. To explain this behavior, several models have been proposed. The simplest scenario involves the presence of at least two environments, in which the indicator is quenched at different rates, often referred to as the multi-site, or for two sites the two-site model \cite{Carraway1991}. In this model, the SV curve is the sum of at least two contributions and written as
\begin{equation}
\frac{I_0}{I}=\bigg( \frac{f_1}{1+K_{SV1} \cdot \left[O_2\right]}+\frac{f_2}{1+K_{SV2}\cdot \left[O_2\right]}\bigg)^{-1}
\label{SVe2}
\end{equation}
where $I_0$ and $I$, respectively, are the luminescence intensities in the absence and presence of oxygen, $f_1$ and $f_2=1-f_1$ are the fractions of the total emission for each component under unquenched conditions, $K_{SV1}$ and $K_{SV2}$ are the associated Stern–Volmer constants for each component, and $\left[O_2\right]$ indicates the oxygen concentration. Since $f_1 + f_2=1$, the following notation will be used in this work: $f_1 = f $ and $f_2 = 1- f $. A simplification of this model is that one of the sites is not quenched and therefore the constant $K_{SV2}$ is zero \cite{Hartmann1996}.
Although this model was introduced for luminescence intensities, it is frequently also used to describe the oxygen dependence of the decay times \cite{Demas1995,Quaranta2012}. Several other more complex models have been proposed \cite{Demas1995,Hartmann1995,Mills1999,Chatni2009} but will not be discussed here.

The luminescence decay time determination is frequently preferable to intensity measurement in sensor implementation because of the proved higher reliability and robustness, since it is not affected by changes in the source intensity or detector sensitivity \cite{Lippitsch1993,Lakowicz2006}. Two approaches can be used to realize decay-time measurement, either using a pulsed excitation (time domain) or modulate the intensity of the excitation (frequency domain). The latter, also known as phase fluorimetry, has the advantage of allowing very simple and low-cost realization and is widely used in commercial applications. Therefore, it was chosen for this work. In this approach, the intensity of the excitation light is modulated. The emitted luminescence light is also modulated but shows a phase shift $\theta$ due to the finite lifetime of the excited state. For a single-exponential decay, the relation between these quantities is
\begin{equation}
\text{tan} \ \theta = \omega \ \tau
\label{theta}
\end{equation}
where $\omega$ is the angular frequency. The range of modulation frequencies to be chosen to determine the intensity decay depends on the lifetimes. The useful modulation frequencies must be high enough so that the phase shift is frequency dependent, but lower than the frequencies where the modulation is not measurable any more.

In the multi-site model, the intensity decay curve is typically no longer a single-exponential decay, even if the decay constant for all sites is the same due, for example, to site-dependent variations of the oxygen diffusion rate \cite{Ogurtsov2001}. A sum of two or more exponentials can well describe the experimental response time of the system to a pulsed excitation \cite{Draxler1995,Demas1995}. In case of a multi-exponential behavior, there is not one single decay time and the relationship between phase shift and decay times must be calculated through the sine and cosine transforms of the intensity decay and the analytical model becomes significantly more complicated \cite{Lippitsch1993,Stehning2004,Digris2005,Lakowicz2006,Ogurtsov2006} . 
Inherent problems of this type of models are from one side, that they may lack relevant physical interpretation when it comes to describing the effect of quenching \cite{Draxler1995}, and from the other, that calculating the oxygen concentration using a non-linear fit procedure to determine all the parameters may become too complex for a robust solution, as required in a sensor.

In most industrial and commercial sensor applications, where the purpose is to calculate the oxygen concentration, it is standard practice to relate the phase shift measured at a single frequency to an apparent or average lifetime using equation (\ref{theta}). Combining equation (\ref{theta}) and assuming the SV relation of equation (\ref{SVe2}) to hold for the decay times, the phase shift and the oxygen concentration results described by the approximate model
\begin{equation}
\frac{\text{tan} \ \theta_0 }{\text{tan} \ \theta} = \bigg( \frac{f}{1+K_{SV1} \cdot \left[O_2\right]}+\frac{1-f}{1+K_{SV2}\cdot \left[O_2\right]} \bigg)^{-1}
\label{theta_full}
\end{equation}
where $\theta_0$ and $\theta$, respectively, are the phase shifts in the absence and presence of oxygen, $f$ and $1-f$ are the fractions of the total emission for each component under unquenched conditions, $K_{SV1}$ and $K_{SV2}$ are the associated Stern–Volmer constants for each component, and $\left[O_2\right]$ indicates the oxygen concentration. The quantities $f$, $K_{SV1}$, and $K_{SV2}$ may result frequency dependent, an artifact of the approximation of the model. Since the purpose of this work is to generate synthetic data to perform the training of the neural network, the model of equation (\ref{theta_full}) is chosen to describe the data, being as simple as possible, and keeping in mind the limited physical meaning.

Another effect which should be included in the model is the temperature dependence of both the unquenched and the quenched lifetimes. As a result, the parameters $\theta_0$, $K_{SV1}$, and $K_{SV2}$ should be characterized by different temperature dependencies.

\section{Experimental Setup}
\label{Experimental}

To determine the parameters for the synthetic data for the training of the neural network and for the validation of the method several luminescence measurements were performed under varying conditions.

The sample used for the characterization and test is a commercially available Pt-TFPP-based oxygen sensor spot (PSt3, PreSens Precision Sensing GmbH, Regensburg, Germany).
To control the temperature of the samples, these were placed in good thermal contact with a copper plate, placed in a thermally insulated chamber. The temperature of this plate was adjusted at a fixed value between 0 $^\circ$C and 45 $^\circ$C using a Peltier element and stabilized with a temperature controller (PTC10, Stanford Research Systems, Sunnyvale, CA USA). The thermally insulated chamber was connected to a self-made gas-mixing apparatus which enabled to vary the oxygen concentration between 0 $\%$ and 20 $\%$ vol $O_2$ by mixing nitrogen and dry air. In the following, the concentration of oxygen will be given in $\%$ of the oxygen concentration of dry air and indicated with $\%$ air. This means, for example, that 20 $\%$ air corresponds to 4 $\%$ vol $O_2$ and 100 $\%$ air corresponds to 20 $\%$ vol $O_2$.  
The absolute error on the oxygen concentration adjusted with the gas mixing device is estimated to be below 1 $\%$ air. 

\begin{figure}[t]
\centering
\includegraphics[keepaspectratio, width=7.5cm]{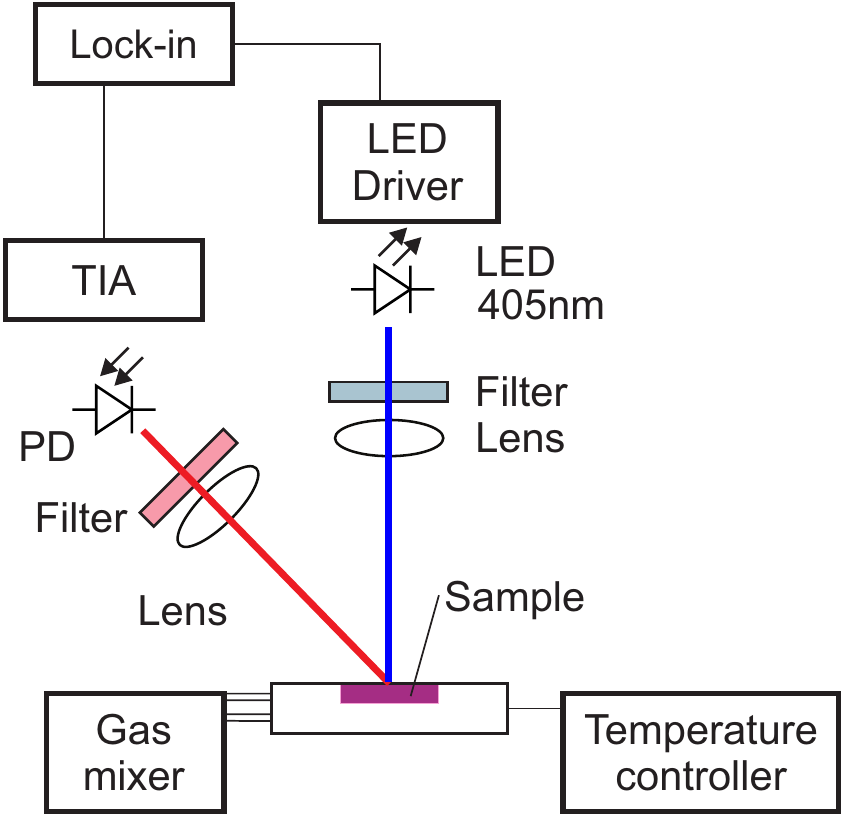}
\caption{Scheme of the optical experimental setup. Blue is the excitation, red the luminescence optical path. PD: photodiode; TIA: trans-impedance amplifier.}
\label{fig:setup}
\end{figure}

The optical setup used in this work for the luminescence measurements is shown schematically in Fig. \ref{fig:setup}.
The excitation light was provided by a 405 nm LED (VAOL-5EUV0T4, VCC Visual Communications Company, LLC, San Marcos, CA USA), filtered by a an OD5 short pass filter with cut-off at 498 nm (Semrock 498 SP Bright Line HC short pass, Semrock, Inc., Rochester, NY, USA) and focused on the surface of the samples with a collimation lens (EO43987, Edmund Optics, Tucson, AZ USA). The luminescence was focussed by a lens (G063020000, LINOS, Qioptiq, Göttingen, Germany) and collected by a photodiode (SFH 213 Osram, Opto Semiconductors GmbH,  Regensburg, Germany).
To suppress stray light and light reflected by the sample surface, the emission channel was equipped with an OD5 long pass filter with cut-off at 594 nm (Semrock 594 LP Edge Basic long pass, Semrock, Inc., Rochester, NY, USA) and an OD5 short pass filter with cut-off at 682 nm (Semrock 682 SP Bright Line HC short pass, Semrock, Inc., Rochester, NY, USA). The driver for the LED and the trans-impedance amplifier (TIA) are self-made.
For the frequency generation and the phase detection a two-phase lock-in amplifier (SR830, Stanford Research Inc.) was used. The modulation frequency was varied between 200 Hz and 20 kHz.

\section{Neural Network Approach}
\label{Neuro}

As described in Section \ref{Theory_Lum}, the intensity decay dependence from the relevant quantities, oxygen concentration, temperature and modulation frequency, is quite complex. The approximated mathematical models described in the literature invariably fail to cover all the details of the measurement setup or sensor. To overcome the limitations of the classical methods, which are based on the theoretical models, this work proposes a new machine learning approach where the mathematical description and the physical significance of the model describing the luminescence decay are irrelevant. With the help of an optimized feed-forward neural network, the sensor learns to relate an input measured quantity to an output quantity from a large number of examples. In other words, the sensor can be considered as a black-box which transforms the input, the phase shift of equation (\ref{theta_full}) measured at several frequencies and a given fixed temperature, into an output, namely the oxygen concentration.

To better describe the method let's introduce the quantity
\begin{equation}
\label{meas}
r (\omega, T, [O_2])\equiv \frac{\tan \theta (\omega, T, [O_2])}{\tan \theta (\omega, T, [O_2]=0) }
\end{equation}
where the meaning of the symbol is the same as described before. 
The method consists in taking a certain (possibly large) number $m$ of measurements of the ratio (\ref{meas}) (this dataset is indicated with $S$) at various known values of the frequency and for a set of values of the oxygen concentration  sampled from a uniform distribution in the range of interest and use it to train a neural network. The proposed approach is a radically different from the commonly used calibration procedures. Usually, the oxygen concentration dependence on the phase shift is programmed in the sensor firmware or in an electronic device as a parametric analytical model. The device-specific parameters are then determined and stored through a calibration. 

To demonstrate the feasibility of the approach, the authors used synthetic data for the training since a sufficiently large number of experimental data could not be acquired at the time of this work. The next step will be the development of a laboratory setup with the capability of acquiring a sufficiently large amount of data under varying conditions, like oxygen concentration, modulation frequency, and temperature.
A further generalization, which would be possible with a larger set of data, is to extend the neural network to give as output both oxygen concentration and temperature. 

In the next subsections, the overview of the method, the generation of the training data and the details of the neural network model are described.

\subsection{Overview of the method}

\begin{figure}[b!]
\centering
\includegraphics[keepaspectratio, width=7.5cm]{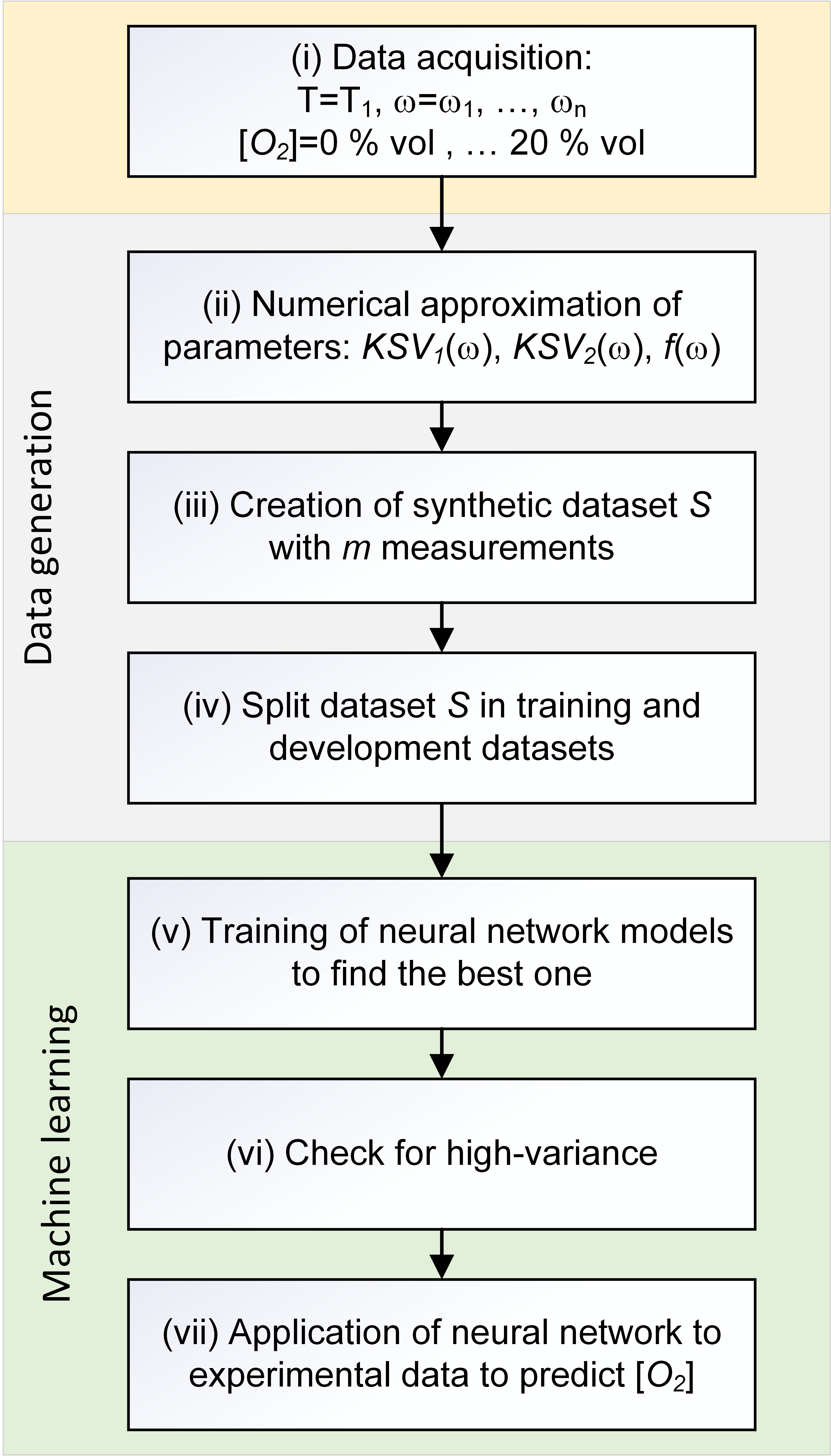}
\caption{Schematic overview of the steps of the machine learning approach.}
\label{fig:flow}
\end{figure}

The schematic overview of the method used is shown in the flowchart of Fig. \ref{fig:flow}. The approach can be divided into the following steps: 
\begin{enumerate}[label=(\roman*)]
\item Data acquisition for different values of frequency, temperature, and oxygen concentration.
\item Determination of a numerical approximation, via interpolation, of the quantities $KSV_1(\omega)$, $KSV_2(\omega)$, and $f(\omega)$ at the chosen temperature $T_1$.
\item Creation of the dataset $S$ with $m$ synthetic measurements using the numerical approximation.
\item Split of the dataset $S$ into a training dataset $S_{train}$ composed of 80 $\%$ of the observations, and a development $S_{dev}$ dataset composed of 20 $\%$ of the observations.
\item Training of several neural network models on the artificial training dataset $S_{train}$.
\item Check for a high-variance (or overfitting) using $S_{train}$ and $S_{dev}$ datasets.
\item Application of the trained neural network model to the experimental dataset to predict the oxygen concentration and comparison with the measured $[O_2]$ quantities.
\end{enumerate}
The steps of the data generation are not essential to the method, but rather are necessary if the availability of the experimental data is limited.

\subsection{Analysis of the raw data and numerical approximation}

The first step of the method (see Fig. \ref{fig:flow}) consists in the acquisition of the data at a given temperature for various modulation frequencies and oxygen concentrations. In this work the ratio defined in (\ref{meas}) was measured at sixteen modulation frequencies, between 500 Hz and 16 kHz, ten values of the oxygen concentration, between 0 $\%$ air and 100 $\%$ air, and five temperatures, between 5 $^\circ$C and 45 $^\circ$C.
The measurements at frequencies lower than 500 Hz were not considered because of the very small value the phase shift of Pt-TFPP assumes. Above 16 kHz the intensity of the modulated light is significantly reduced, which results in higher noise in the phase. Therefore those frequencies were also not used in this study.

\begin{figure}[htb]
\centering
\includegraphics[keepaspectratio, width=9cm]{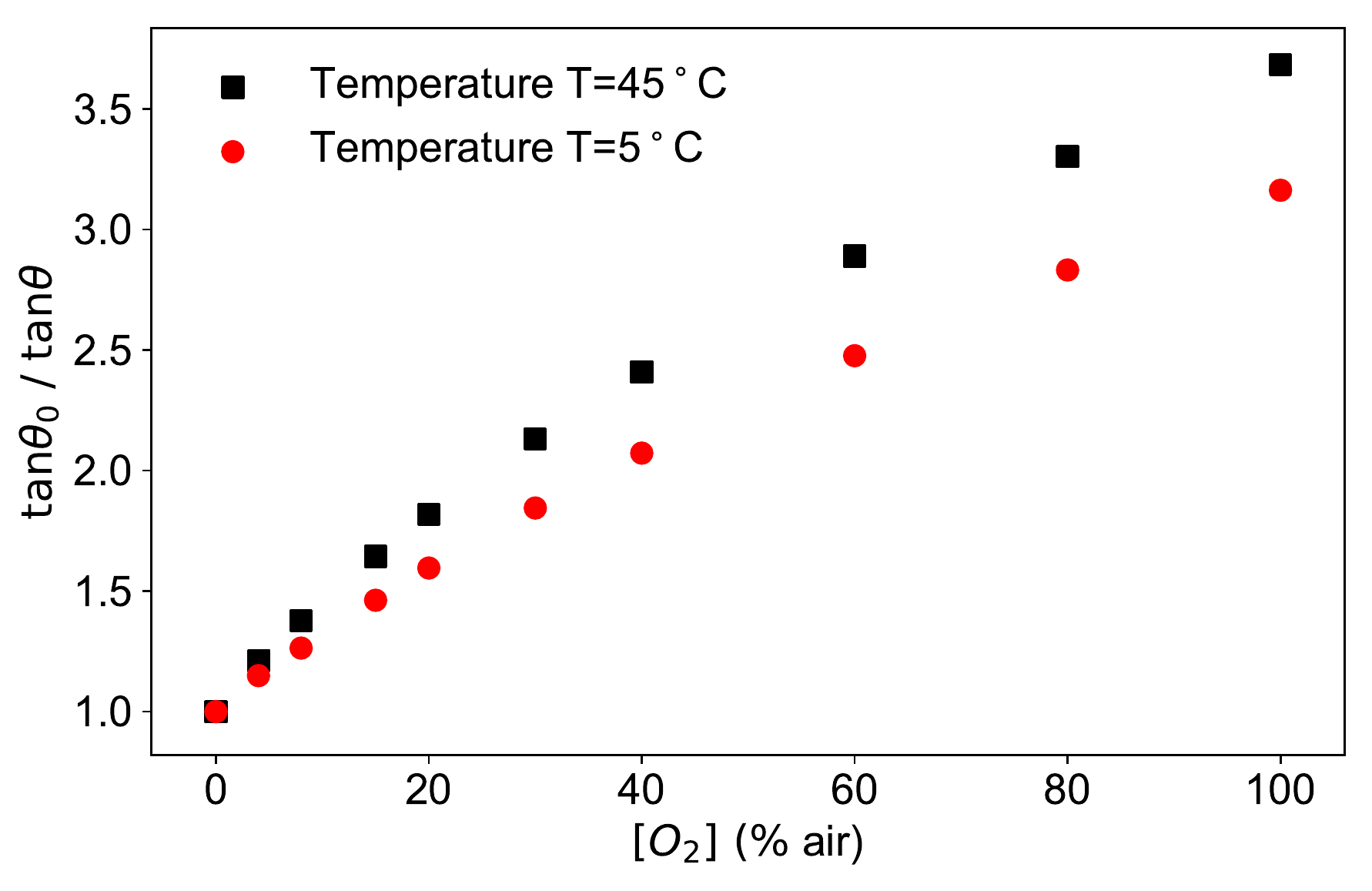}
\caption{Phase shift measured for modulation frequency $2\pi \omega=6$kHz, at two different temperatures 5 $^\circ$C and 45 $^\circ$C as a function of the oxygen concentration.}
\label{fig:raw_o2}
\end{figure}

As an example, the phase shifts measured at a fixed modulation frequency of 6 kHz as a function of the oxygen concentration are plotted as $\tan \theta_0 / \tan \theta$ in Fig. \ref{fig:raw_o2} for two temperatures.
The figure shows the typical dependence of the phase shift on two of the parameters at disposal, oxygen concentration and temperature.

\begin{figure}[hbt]
\centering
\includegraphics[keepaspectratio, width=9cm]{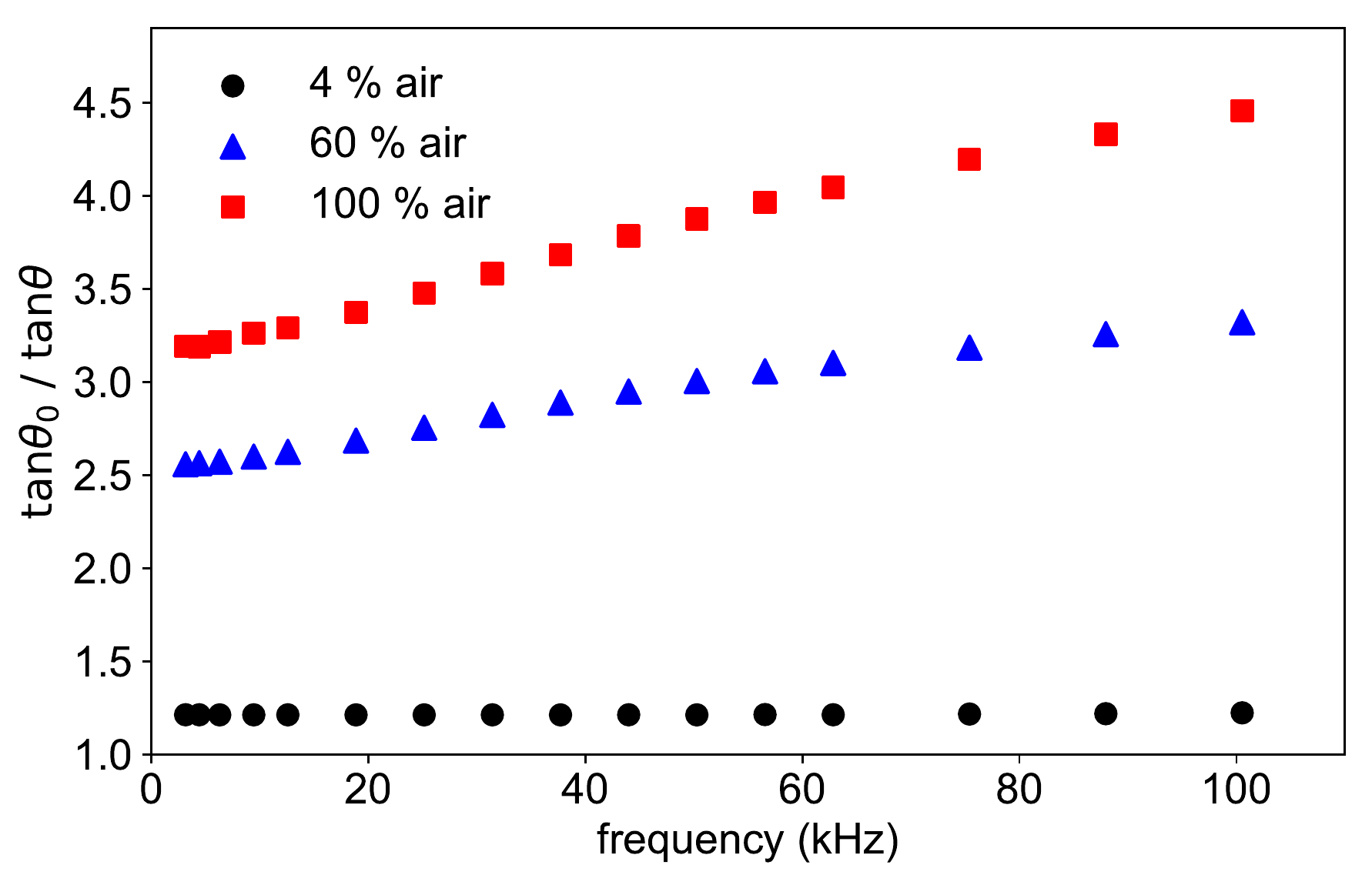}
\caption{Phase shift measured at three different oxygen concentrations at the temperature T=45 $^\circ$C as a function of the modulation frequency.}
\label{fig:raw_f}
\end{figure}

The frequency dependence of the phase shift is shown in Fig. \ref{fig:raw_f} for three different concentrations for a fixed temperature of 45 $^\circ$C. As can be seen from the figure, the ratio $\tan \theta_0 / \tan \theta$ shows a frequency dependence, which is stronger the higher the concentration is. This dependence is due to the oversimplification of equation (\ref{theta}), which is approximately correct in the absence of oxygen, and therefore without quenching, but does not hold for higher concentrations.

The second step of the method (see Fig. \ref{fig:flow}) consists in the determination of the best numerical approximation of the parameters of the theoretical model. This step is required to generate a large number of data for the training of the neural network. Ideally, as mentioned above, this step should be replaced by a large number of experimental data, which were not available at the moment of this work. It is also to be noticed, that the physical meaning of equations and parameters used to generate the synthetic data is not relevant and, therefore, will not be discussed here.

For this purpose a numeric approximation for the functions $f(\omega, T)$, $KSV_1(\omega, T)$, and $KSV_2(\omega,T)$ was determined by fitting the experimental data according to the model of equation (\ref{theta_full}) via standard non-linear fitting procedures. For simplicity, in this work the temperature was kept constant during the analysis. The data shown here correspond to $T_1=45 \ ^\circ$C.

\subsection{Generation of artificial training data}

\begin{figure}[htb]
\centering
\includegraphics[keepaspectratio, width=15.5cm]{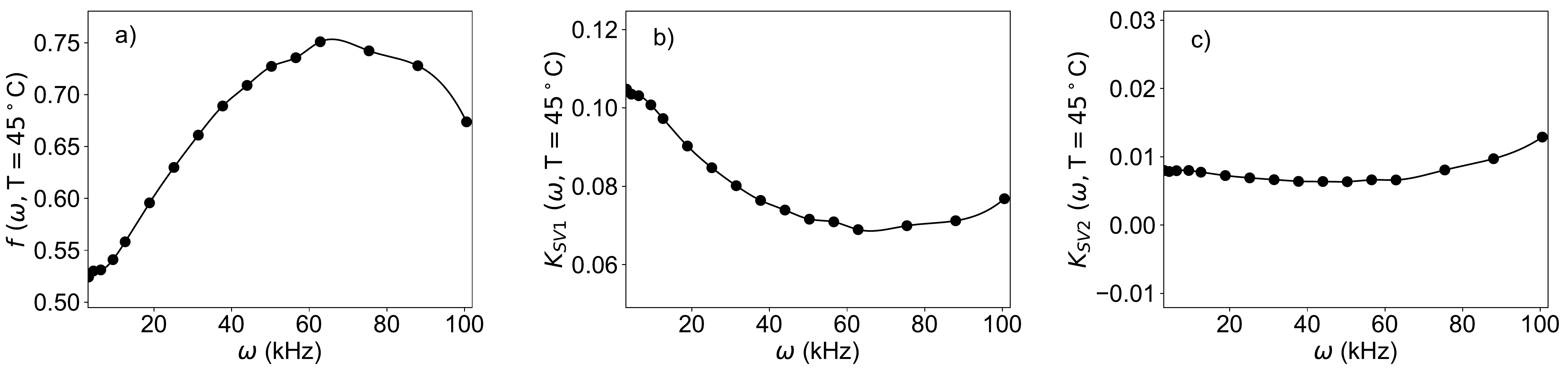}
\caption{Dependence of the parameters of equation (\ref{theta_full}) from the angular frequency of the modulation. Dots: result of the fit of the experimental data; solid line: spline approximation used to generate the synthetic data.}
\label{fig:fit_model}
\end{figure}

After the determination of the numerical approximation of the parameters, the equation (\ref{theta_full}) was used to generate a large number of synthetic data for $r (\omega, T, [O_2])$ which are needed for the training of the neural network (step (iii) in Fig. \ref{fig:flow}).
To facilitate the implementation in the programming language Python$\texttrademark$, it is advantageous for the parameters $f(\omega, T)$, $KSV_1(\omega, T)$, and $KSV_2(\omega,T)$ to be functions defined on a continuous domain rather than on a discrete number (sixteen) of modulation frequencies. For this purpose, a spline of the third order using the function {\sl interp1d} from the package {\sl scipy} \cite{scipy2018} of the programming language Python was implemented.
The frequency dependence of the parameters $f(\omega, T)$, $KSV_1(\omega, T)$, and $KSV_2(\omega,T)$, as well as the spline used in the code, are shown in Fig. \ref{fig:fit_model}. 

The goal of the network is to predict the oxygen concentration from an array of values of $r (\omega, T, [O_2])$ evaluated at a discrete set of sixteen $\omega_i$, with $i=1,...16$, that have been used for the measurements. Each array ${\bf r} = (r_1, r_2, ..., r_{16})$ with $r_i = r (\omega_i, T, [O_2]_j)$ and $i=1,...16$ will be called in this paper an observation. Each observation will be indicated by a superscript $[j]$. So $r_i^{[j]}$ indicates  $r = r (\omega_i, T, [O_2]_j)$ and ${\bf r}^{[j]} = (r_1^{[j]}, r_2^{[j]}, ..., r_{16}^{[j]}) = (r (\omega_1, T, [O_2]_j), ..., r (\omega_{16}, T, [O_2]_j))$. An observation corresponds to a specific value of the oxygen concentration.

The synthetic data consist of a set $S$ of $m=5000$ observations using oxygen concentration values uniformly distributed between 0 $\%$ air and 110 $\%$ air. 
The value of 110 $\%$ air is chosen since the neural network predictions tend to be less good when dealing with observations that are close to a boundary. The tests show that, having only observations for the training with $[O_2] < 100 \  \%$ air makes the network predictions less accurate in the prediction for $[O_2]$ close to 100 $\%$ air. A discussion of this effect can be found in \cite{Michelucci2018_2}.

Next, (step (iv) in Fig. \ref{fig:flow}) these data are split randomly in a training dataset $S_{train}$ containing 80 $\%$ of the data, so 4000 observations, used to train the network, and a development dataset $S_{dev}$ containing 20 $\%$ of the data, so 1000 observations, used to test the generalization of the network when applied to unseen data. For the validation, the neural network model is applied to a test dataset, indicated with $S_{test}$, whose observations are the ten experimental measurements. Finally, the predictions of the oxygen concentration are compared to the measured values.

\subsection{Neural network model}

\begin{figure}[hbt]
\centering
\includegraphics[keepaspectratio, width=7.5cm]{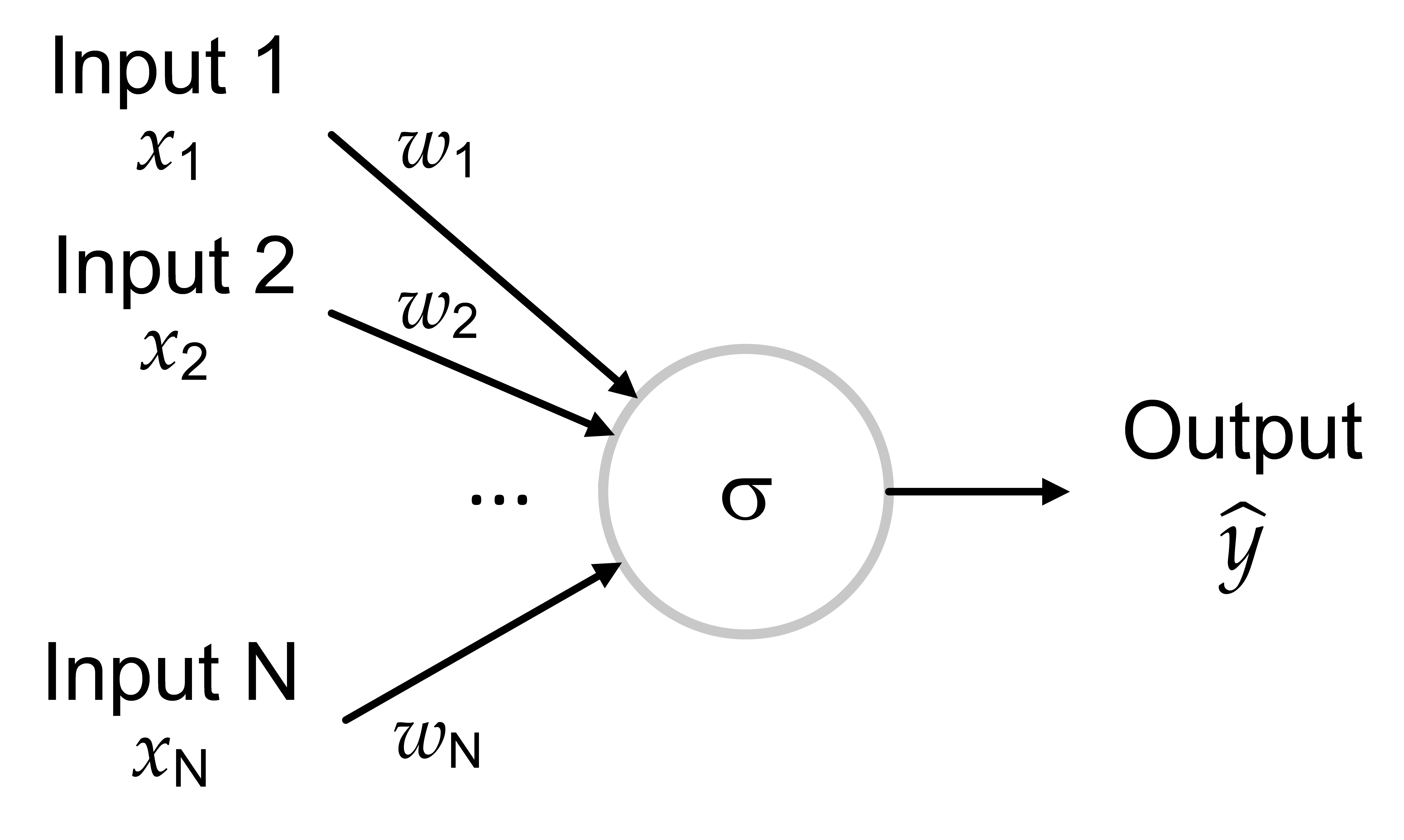}
\caption{A schematically depicted neuron. It applies a non-linear transformation to the inputs with the sigmoid activation function to obtain its output.
}
\label{fig:Neuron}
\end{figure}

The true machine learning (steps (v-vii) in Fig. \ref{fig:flow}) starts with the  neural network model. The building block of the network used in this work is a neuron, which transforms a set of real numbers given as inputs $x_i$ in an output $\hat y$ using the formula
\begin{equation}
\label{neuron_f}
\hat y = \sigma \left( \sum_{\text{number of inputs}} w_i x_i + b \right)
\end{equation}
where $w_i$ are called weights, $b$ bias, and $\sigma$, which is called the activation function, is the sigmoid function that has the analytical form 
\begin{equation}
\sigma(z) = \frac{1}{1+e^{-z}}.
\end{equation}
This is schematically depicted in Fig. \ref{fig:Neuron}.

\begin{figure}[hbt]
\centering
\includegraphics[keepaspectratio, width=12cm]{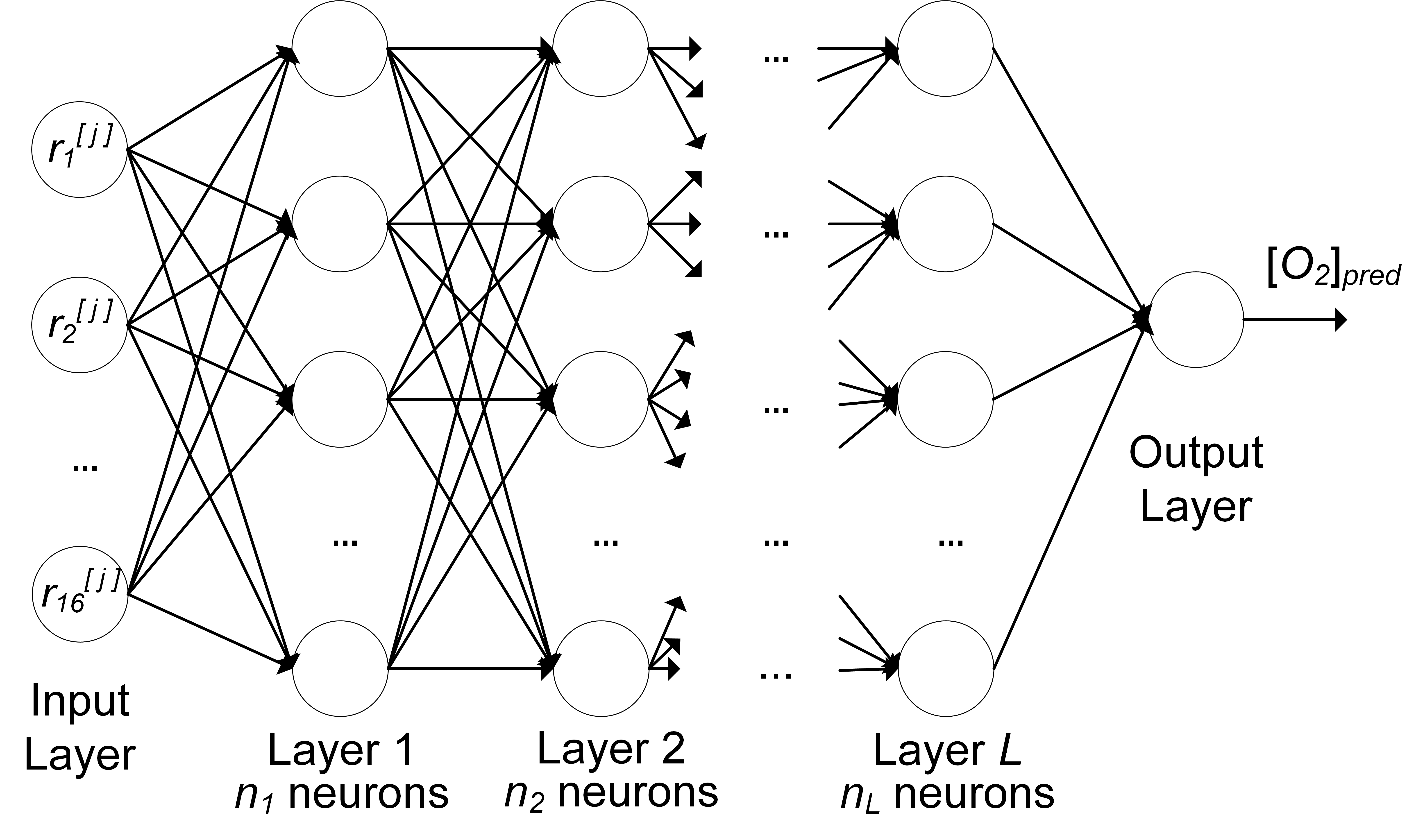}
\caption{Architecture of the feed-forward neural network with $L$ layers, each having a number of neurons $n_i$. $r_i^{[j]}$ is the $i$-th feature of the $[j]$-th observation; the output is the predicted oxygen concentration $[O_2]_{pred}$.}
\label{fig:NN}
\end{figure}

The architecture of the neural network of this work is shown schematically in Fig \ref{fig:NN}.
The network includes a number of layers $L$, each the same number of neurons $n_i$. The architecture of Fig. \ref{fig:NN} is of the type feed-forward, where each neuron in each layer gets as input the output of all neurons in the previous layer before, and feeds its output to each neuron in the subsequent layer.

A neural network model is made of three parts: the network architecture, described above (Fig. \ref{fig:NN}), the cost function $J$, and an optimizer. Training the network means finding the best weights and bias of all neurons of the network (cf. equation (\ref{neuron_f}) for a single neuron) to minimize $J$. The optimizer is the algorithm used to minimize the cost function. Since it is a regression problem, the cost function $J$ is taken to be the mean squared error, defined as the squared average of the absolute value of the difference between the predicted oxygen concentration values and the expected ones. To minimize the cost function the optimizer Adaptive Moment Estimation (Adam) \cite{Kingma2014} was used. The implementation was performed using the TensorFlow$\texttrademark$ library.

To study the dependence of the results from the architecture of the network, a process called  hyperparameter optimization, both the number of layers $L$ and the number of neurons per layer  $n_i$ were varied. 
For all the neurons the sigmoid function was taken as activation function. For the output neuron, that has as output the predicted values $[O_2]_{pred}$, the function $110 \cdot \sigma$ was chosen, since the dataset contains training observations with $[O_2]$ up to 110 $\%$ air.
The predictions of the network are then compared to identify the best architecture.

The metric used to compare results from different network models is the mean absolute error (MAE), defined as the average of the absolute value of the difference between the predicted and the expected or measured oxygen concentration.  For example, for the $S_{test}$ dataset 
\begin{equation}
\label{MAE}
\text{MAE}(S_{test}) = \frac{1}{|S_{test}|} \sum_{S_{test}}|[O_2]_{pred}-[O_2]_{meas}|
\end{equation}
with $|S_{test}|$ the size (or cardinality) of the dataset $S_{test}$. 
The further quantity used to analyze the performance of the network is the 
absolute error (AE) of a given observation $j$, defined as the absolute value of the difference between the predicted and the measured $[O_2]^{[j]}$ value 
\begin{equation}
\label{AE}
\text{AE}^{[j]} = |[O_2]^{[j]}_{pred}-[O_2]^{[j]}_{meas}|.
\end{equation}

\section{Results and discussion}
\label{Results}

\begin{figure}[htb]
\centering
\includegraphics[keepaspectratio, width=9.5cm]{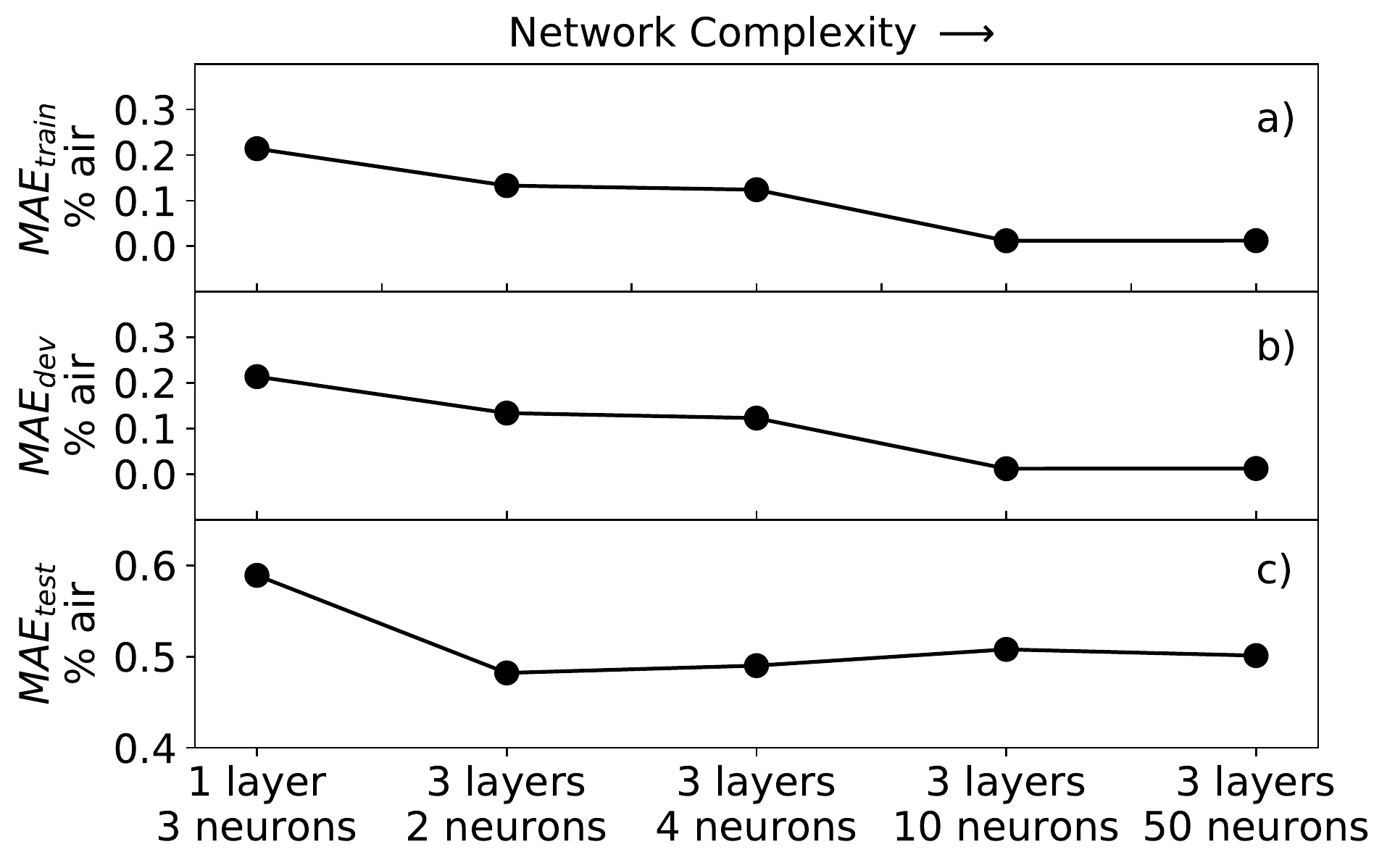}
\caption{Mean absolute error of the network applied to different datasets for various networks architectures: a) MAE$_{train}$ for training dataset, b) MAE$_{dev}$ for development dataset and c) MAE$_{test}$ for the test dataset. The complexity of the network increases left to right.}
\label{fig:results}
\end{figure}

To investigate the performance of the neural network the architecture was varied by changing the number of layers, between 1 and 3, and the number of neurons per layer, between 3 and 50. The training was performed for all the trials with batch gradient descent for $10^5$ epochs and learning rate of 0.001. The latter was chosen because of the fast convergence of the cost function. For each trial the MEA of equation (\ref{MAE}) was calculated.
The result of the analysis is summarized in Fig. \ref{fig:results}.

As expected, the values of all three MAE are larger for neural networks with lower effective complexity, on the left of the plot, then for ones with higher complexity, on the right of the plot, regardless of the dataset to which the network is applied. This result reflects the fact that, if the network is not complex enough, it does not perform well since it cannot learn the subtleties of the data. 
From Fig. \ref{fig:results}, a) and Fig. \ref{fig:results}, b) can be noted that the MAE$_{train}$ and MAE$_{dev}$ have similar behavior and assume 
almost the same values, going to zero for increasing complexity. For the architectures studied in this work, both MAE$_{train}$ and MAE$_{dev}$ reach a minimum of 0.012 $\%$ air for the network with 3 layers and 50 neurons. The reason for the error going almost to zero is that both the training and development datasets contained synthetic data generated by the same function, therefore without any experimental noise. 
Typically when training neural network models, it is important to check if we are in a so-called overfitting regime. The essence of overfitting is to have unknowingly extracted some of the residual variation (i.e., the noise or errors) as if that variation represented an underlying model structure \cite{Burnham}. In this work, with increasing complexity, the network will never go into such a  regime, since the development dataset is a perfect representation of the training dataset. This leads to almost identical MAE$_{train}$ and MAE$_{dev}$ error values, regardless of the network architecture effective complexity.

The MAE$_{test}$ (Fig. \ref{fig:results}, c), on the other hand, shows a different behavior, rapidly improving from the simplest architecture of 1 layer and 3 neurons, but then not further decreasing by increasing the complexity of the neural network. In other words, as soon as the complexity of the network is enough to  approximate well the inverse function of equation (\ref{theta_full}), the MAE$_{test}$ stabilizes at a value of approximately 0.5 $\%$ air.

The  reason why MAE$_{test}$ does not go to zero is twofold.
First, $S_{test}$ includes the experimental measurements, which are affected by an experimental error. Since the neural network was trained with synthetic data, it does not include such an error and will consequently always have a certain deviation in predicting $[O_2]$.

\begin{figure}[b!]
\centering
\includegraphics[keepaspectratio, width=9cm]{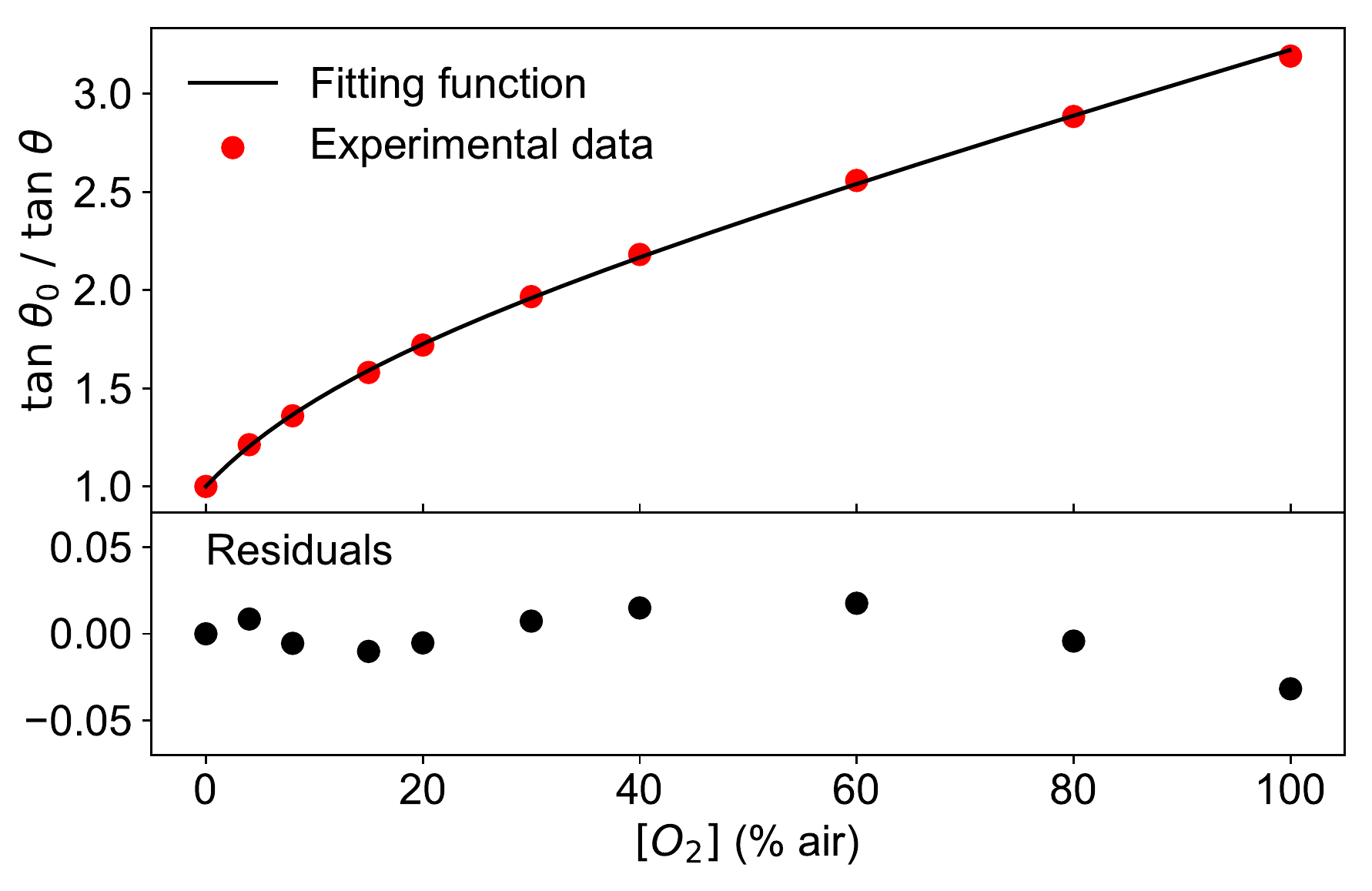}
\caption{Upper panel: phase shift measured at a modulation frequency of 6 kHz and at the temperature of 45 $^\circ$C for various values of oxygen concentration (circles). The line is the fit obtained with the function (\ref{theta_full}). Lower panel: Residuals of the fit.}
\label{fig:fit_residuals}
\end{figure}

Second and most importantly, the theoretical model of equation (\ref{theta_full}) does not approximate the experimental data sufficiently well, as is illustrated in Fig. \ref{fig:fit_residuals}. In the upper panel, $\tan \theta_0 / \tan \theta$ measured at a modulation frequency of 6 kHz and the temperature of 45 $^\circ$C is shown as a function of the oxygen concentration together with the best fit using equation (\ref{theta_full}). The residuals, calculated as the difference between the fit and the measured data, are plotted in the lower panel of Fig. \ref{fig:fit_residuals} and show that the deviation of the measurement and model increases with higher oxygen concentration.
Therefore, the training performed with synthetic data has the disadvantage to lead to a network which learns from a dataset with different functional shape (albeit not extremely so) than the test dataset. This contribution to the MAE$_{test}$ could be eliminated using experimental measures as a training dataset, allowing thus to achieve even better predictions of the oxygen concentration than 0.5 $\%$ air.

\begin{figure}[htb]
\centering
\includegraphics[keepaspectratio, width=9cm]{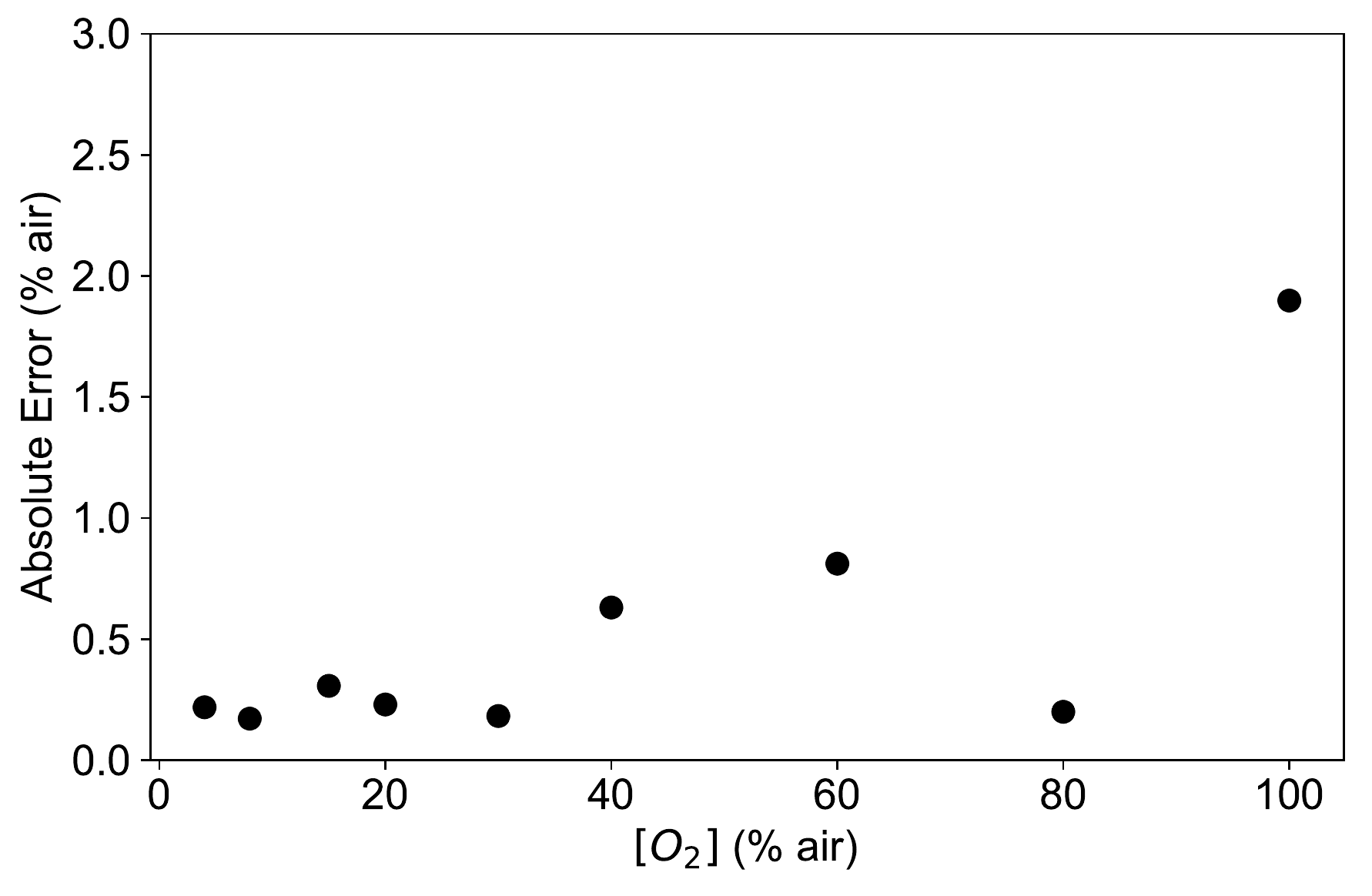}
\caption{Absolute error (AE) of the neural network model prediction applied to the experimental data at the temperatures 45 $^\circ$C for a network with 3 layers and 10 neurons.}
\label{fig:error_vs_O2}
\end{figure}

These results are supported by the analysis the dependence of the absolute error AE on the oxygen concentration. As an example, the absolute error calculated as in equation (\ref{AE}) is shown in Fig. \ref{fig:error_vs_O2} for a network with 3 layers and 10 neurons. As it can be seen from Fig. \ref{fig:error_vs_O2}, the AE is below 0.5 $\%$ air for low oxygen concentrations, tends to increase with higher $[O_2]$ values and reaches its maximal value of 2 $\%$ air at 100 $\%$ air. This is consistent with the deviations shown in Fig. \ref{fig:fit_residuals}, indicating that the fitting function works worst at 100 $\%$ air.

\begin{figure}[hbt!]
\centering
\includegraphics[keepaspectratio, width=9cm]{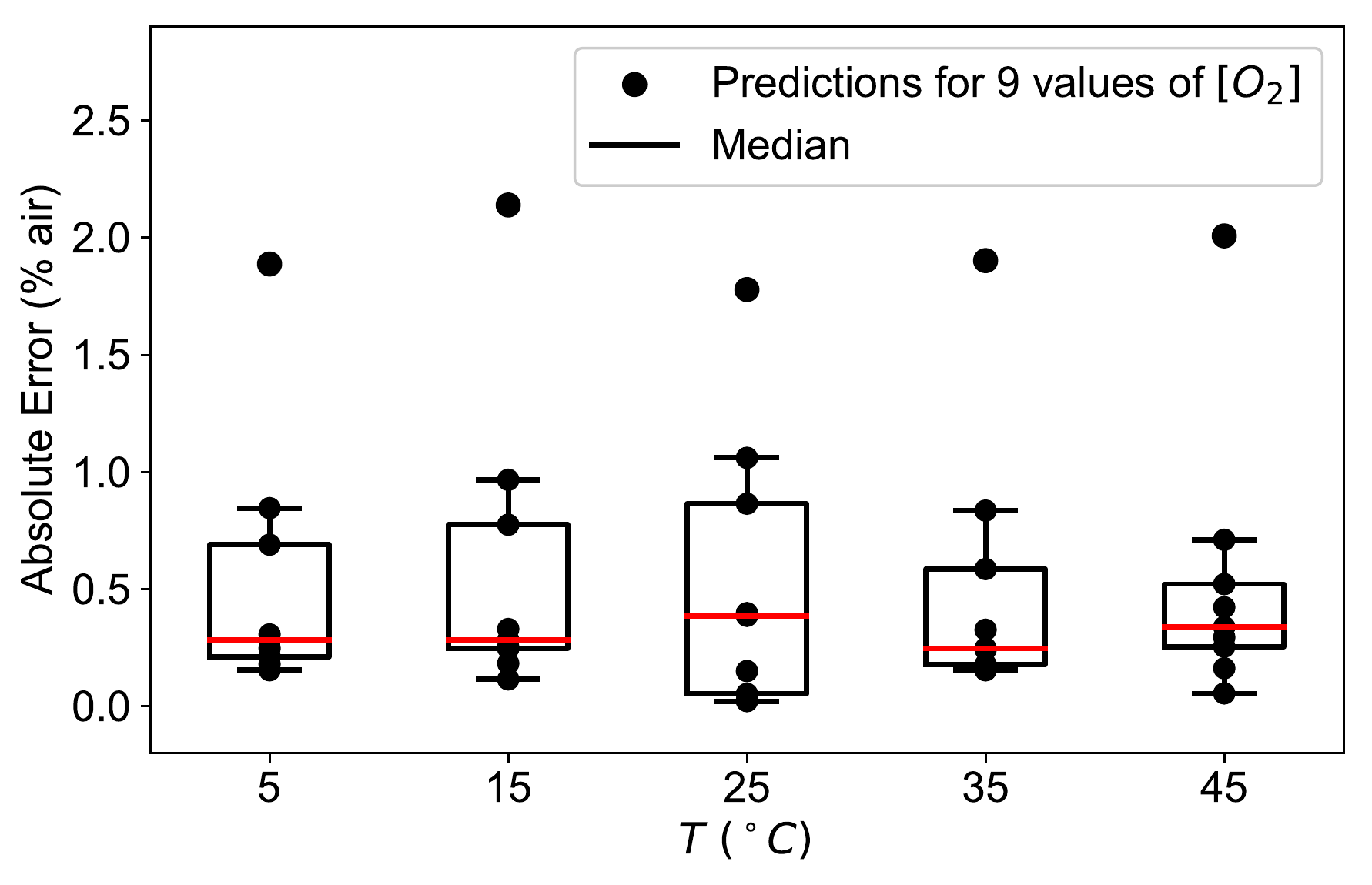}
\caption{Absolute error (AE) distribution of the neural network model prediction applied to the experimental data for different concentrations calculated at the temperatures of  5 $^\circ$C, 15 $^\circ$C, 25 $^\circ$C, 35 $^\circ$C, and 45 $^\circ$C for a network with 3 layers and 10 neurons.}
\label{fig:box_plot_T}
\end{figure}

The above mentioned observations are valid at all the temperature studied, as shown in Fig. \ref{fig:box_plot_T}. For each temperature the absolute error is calculated for the available oxygen concentrations and displayed as a box plot, where the median is visible as a red line.
For all the temperature investigated the median remains below 0.5 $\%$ air for low oxygen concentrations and the absolute error has a maximal value of 2 $\%$ air at 100 $\%$ air, plotted separately in Fig. \ref{fig:box_plot_T} for clarity. 

From a theoretical point of view a network with one layer can approximate any non-linear function \cite{Montufar2014,Fortuner2017}. However, a network with more layers but fewer neurons in total may be able to capture specific features of the data better and faster, that is with a smaller number of epochs in the training. The analysis of the different network architectures performed in this work supports these results. To reach the same value for the MAE with networks with one layer requires 5 to 10 times more epochs for the training. Consequently, it is much more efficient to choose a network with more layers.
In this work, networks with architectures more complex than with 3 layers and 50 neurons were also studied. Since the MAE$_{test}$ stabilizes already for simple architectures, as shown in Fig. \ref{fig:results}, an increase in the complexity will not further improve the performance.

\section{Conclusions}

This work explored a new approach to optical luminescence sensing proving that to build an accurate oxygen sensor is not necessary to implement complex non-linear pseudo-physical models to describe the dependence of the measured quantity, the phase shift, from the oxygen concentration. The sensor-specific deviations from a simple SV model may be caused, for example, by the method for the immobilization of the indicators in the substrate, or by luminescence from components typically included in a sensor, like absorption filters or glues. 
Therefore, to reach a high accuracy, the classical approach requires empirically modeling the dependence of the parameters $f$, $K_{SV1}$, $K_{SV2}$ of the inverted SV multi-site model from all the relevant influencing quantities, like for example the temperature or the modulation frequency. The resulting algorithms frequently are computationally intensive and per-definition only an approximation. Furthermore, for a commercial solution, it is desirable for the chosen parametrization to be valid for all the sensors, with only a few parameters which need to be determined during a device-specific calibration. This imposes higher requirements, a therefore costs, on the device components, and may require further compromising on the accuracy. The proposed artificial intelligence approach has the potential to overcome these limitations.

Several neural network architectures were tested, demonstrating that already networks with rather simple structures can predict the oxygen concentration with a mean absolute error MAE$_{test}$ of 0.5 $\%$ air. By an analysis of discrepancies as a function of the oxygen concentration was possible to identify the main contribution to the error. The results show that the absolute error AE increases with increasing concentrations, going from well below 0.5 \% air at 30 \% air to a maximum of 2 \% air at 100 \% air. The main contribution to the AE was identified in the poor agreement of the conventional model describing the quenching of the luminescence, which was used to generate the training data.  By performing the training on experimental data this error is expected to decrease significantly. 

This work paves the road to a new and completely different approach in sensor development. Using artificial intelligence is not necessary to define any models with parameters to capture all influencing factors. Once the sensor hardware is given, that is all the optical, electronic, mechanical and chemical components are assembled, a neural network can be trained to learn to predict one, or possibly more, quantities of interests, in this work the oxygen concentration. The advantages of this approach will impact positively, for example, on the scale-up in sensor fabrication because the requirements on the hardware can be relaxed. The device-specific characteristics will be accounted for by the trained neural network.

\vspace{6pt}

\authorcontributions{U. Michelucci analyzed the data, developed the neural network and wrote the paper. M. Baumgartner built the laboratory setup and performed the experiments, F. Venturini conceived the study, analyzed the data and wrote the paper.}


\conflictsofinterest{The authors declare no conflict of interest.} 

\abbreviations{The following abbreviations are used in this manuscript:\\

\noindent 
\begin{tabular}{@{}ll}
SV & Stern-Volmer\\
LED & Light-emitting diode\\
TIA & trans-impedance amplifier \\
Adam & Adaptive Moment Estimation \\
MAE & Mean absolute error\\
AE & Absolute error
\end{tabular}}

\appendixtitles{no} 
\appendixsections{multiple} 




\reftitle{References}

\bibliographystyle{unsrtnat}





\end{document}